# A scalable approach for tree segmentation within small-footprint airborne LiDAR data

Hamid Hamraz[a*], Marco A. Contreras[b], Jun Zhang[a]

a: Department of Computer Science, b: Department of Forestry

University of Kentucky, Lexington, KY 40506, USA

hhamraz@cs.uky.edu, marco.contreras@uky.edu, jzhang@cs.uky.edu

* Corresponding Author:    hhamraz@cs.uky.edu  +1 (859) 489 1261

**Abstract.**

This paper presents a distributed approach that scales up to segment tree crowns within a LiDAR point cloud representing an arbitrarily large forested area. The approach uses a single-processor tree segmentation algorithm as a building block in order to process the data delivered in the shape of tiles in parallel. The distributed processing is performed in a master-slave manner, in which the master maintains the global map of the tiles and coordinates the slaves that segment tree crowns within and across the boundaries of the tiles. A minimal bias was introduced to the number of detected trees because of trees lying across the tile boundaries, which was quantified and adjusted for. Theoretical and experimental analyses of the runtime of the approach revealed a near linear speedup. The estimated number of trees categorized by crown class and the associated error margins as well as the height distribution of the detected trees aligned well with field estimations, verifying that the distributed approach works correctly. The approach enables providing information of individual tree locations and point cloud segments for a forest-level area in a timely manner, which can be used to create detailed remotely sensed forest inventories. Although the approach was presented for tree segmentation within LiDAR point clouds, the idea can also be generalized to scale up processing other big spatial datasets.

# 1 Introduction

Individual tree information is increasingly becoming the preferred data precision level to accurately and efficiently monitor, assess, and manage forest and natural resources (Chen *et al.*, 2006; Koch *et al.*, 2006; Schardt *et al.*, 2002). In the last two decades, airborne light detection and ranging (LiDAR) technology has brought drastic changes to forest data acquisition and management by providing inventory data at unprecedented spatial and temporal resolutions (Ackermann, 1999; Maltamo *et al.*, 2014; Shao and Reynolds, 2006; Swatantran *et al.*, 2016; Wehr and Lohr, 1999). However, to obtain accurate tree level attributes such as crown width and tree height as well as derivative estimates such as diameter at breast height (DBH), volume, and biomass, accurate and automated tree segmentation approaches are required (Schardt *et al.*, 2002).

Numerous methods for tree segmentation within LiDAR data have been proposed (Duncanson *et al.*, 2014; Hamraz *et al.*, 2016; Hu *et al.*, 2014; Jing *et al.*, 2012; Li *et al.*, 2012; Persson *et al.*, 2002; Popescu and Wynne, 2004; Véga and Durrieu, 2011; Véga *et al.*, 2014; Wang *et al.*, 2008). Nevertheless, these methods have only been experimented for small forested areas and none of them have thoroughly considered scalability; LiDAR data covering an entire forest is much more voluminous than the memory of a typical workstation and may also take an unacceptably long time to be sequentially processed. Also, given the continuous advancements of the sensor technology (Swatantran *et al.*, 2016), the LiDAR point clouds will be acquired with less costs and greater resolutions, which in turn increases the need for more efficient and scalable processing schemes.

A few studies have considered processing LiDAR data (Thiemann *et al.*, 2013; Zhou and Neumann, 2009) using streaming algorithms (Pajarola, 2005), where the spatial locality of the LiDAR data is used to construct out-of-core algorithms. However, streaming algorithms are unable to reduce the time required for processing because of their inherently sequential processing scheme. A number of recent studies have considered leveraging the power of multicore and/or GPU (shared memory) platforms for processing LiDAR data for efficient DEM modeling (Guan and Wu, 2010; Oryspayev *et al.*, 2012; Sten *et al.*, 2016; Wu *et al.*, 2011), or for 3D visualization (Bernardin *et al.*, 2011; Li *et al.*, 2013; Mateo Lázaro *et al.*, 2014), although shared-memory platforms are also bounded in the amount of memory and the number of processing units.

On the other hand, processing geospatial data such as LiDAR data can be parallelized by partitioning the data into tiles (commonly used for data delivery purposes) and distributing the tiles to different processors on a distributed architecture. Huang *et al.* (2011) proposed a master-slave distributed method for parallelizing inverse distance weighting interpolation algorithm. Guan *et al.* (2013) designed a cloud-based process virtualization platform to process vast quantities of LiDAR data. Barnes (2016) parallelized Priority-Flood depression-filling algorithm by subdividing a DEM into tiles. However, the above distributed approaches were designed and used for perfectly parallel problems while, in case of non-perfectly parallel problems, dealing with the data near the boundaries of the tiles is not trivial and should be elaborated according to the specifics of the application (Werder and Krüger, 2009).

Accounting for the data near the tile boundaries, a distributed density-based clustering for spatial data (Ester *et al.*, 1996) was presented by Xu et al. (2002). The authors proposed a master-slave scheme in which the master spawns a number of slaves to perform the clustering and return the result back to the master, who then combines the results. The scheme relies on a data placement strategy for load balancing in which the master partitions the data and distributes the portions among the slaves for processing, hence the runtime is determined by the last slave that finishes its job. Distributing the data and merging the results by the master are also sequential procedures and may yield performance bottlenecks. A more recent work (He *et al.*, 2011) has presented a version of the density-based clustering tailored to run on a MapReduce infrastructure (Dean and Ghemawat, 2008) performing four stages of MapReduce for indexing, clustering, as well as identifying and merging boundary data. The MapReduce infrastructure, although constraining the programming model, has the advantage of built-in simplicity, scalability, and fault tolerance. Thiemann et al. (2013) have presented a framework for distributed processing of geospatial data, where partitioning the data to tiles with overlapping areas near the borders is their core solution. The overlapping area should be at least as big as the required neighborhood for processing a local entity and the produced overlapping result may require special treatment to be unified. The authors used the map phase of the Hadoop MapReduce infrastructure (White, 2012) for clustering buildings of large urban areas and the overlapping result was unified separately afterwards.

Although there are various methods proposed for tree segmentation, only few studies have considered scalable processing of large geospatial data – there is specifically no study considering forest-level datasets. This is increasingly important when obtaining tree-level information for areas other than small-scale plots, which is often the case when obtaining LiDAR data. This paper presents and analyzes a distributed approach that accounts for the data near the tile boundaries and uses a tree segmentation algorithm as a building block in order to efficiently segment trees from LiDAR point clouds representing an entire forest. For experimentation, the approach was implemented using message passing interface (MPI) (Walker, 1994).

## 2 Materials and methods

### 2.1 LiDAR data

We used LiDAR data acquired over the University of Kentucky Robinson Forest (Lat. 37.4611, Long. -83.1555), which covers an aggregated area of 7,441.5 ha in the rugged eastern section of the Cumberland Plateau region of southeastern Kentucky in Breathitt, Perry, and Knott counties (37°28′23″N 83°08′36″W) (Overstreet, 1984). The LiDAR data is a combination of two datasets collected with the same LiDAR system (Leica ALS60 at 200 kHz flown with an average speed of 105 knots) by the same vendor. One dataset was low density (~1.5 pt/m$^2$) collected in the spring of 2013 during leaf-off season (average altitude of 3,096 m above the ground). The second dataset was high density (~25 pt/m$^2$) collected in the summer of 2013 during leaf-on season (average altitude of 196 m above the ground). The combined dataset has a nominal pulse spacing (NPS) of 0.2 m and was delivered in 801 square (304.8 m side ~ 9.3 ha area) tiles (Figure 1), each containing about 5 million LiDAR points on average and occupying about 400 MB of disk space. The entire LiDAR dataset contains over 4 billion points and occupies 320 GB of disk space.

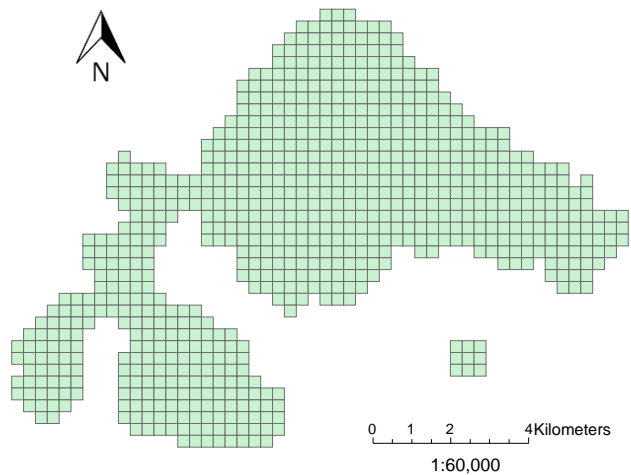

Figure 1. LiDAR tile map of Robinson Forest consisting of 801 9.3-ha tiles.

## 2.2 Distributed processing

In a distributed processing environment, the LiDAR data representing tree crowns located across tile boundaries is split into two or more pieces that are processed by different processing units. Identifying such crown pieces, unifying them, and efficiently managing the distributed resources to run with a reasonable speedup are the main challenges of a distributed approach. We propose a master-slave distributed approach, where the master is in charge of maintaining the global tile map and coordinating how to process individual tiles and their boundary data while the slaves perform the actual tree segmentation.

Tile boundary data (solid/striped colored regions in Figure 2) likely represent tree crowns located between two tiles (light-colored) – hereafter referred to as edge data – or among three or four tiles (dark-colored) – hereafter referred to as corner data. After segmenting a tile, all segmented crowns that have at least one LiDAR point within a horizontal distance of 2×NPS from a tile edge form part of the boundary data. The crowns that are adjacent to only one edge (solid light colored) are regarded as a part of the associated edge data and those that are adjacent to exactly two edges (solid dark colored) are regarded as a part of the associated corner data.

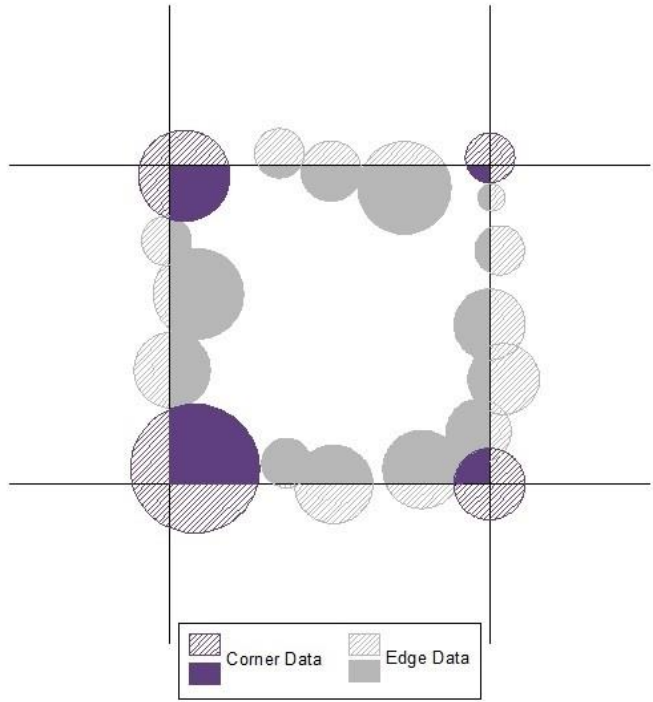

Figure 2 A schematic of a tile with the two types of boundary data. The solid-colored tree crown pieces inside the tile should be unified with the corresponding stripe-colored parts outside.

Figure 3 and 4 show the flowcharts of the master and the slave processes. It is assumed that all processes can independently input tiles data and output results. Such an assumption can reasonably be fulfilled by using a supercomputing infrastructure with a unified file system (typically designed to efficiently support all existent physical processing cores), by maintaining the tiles and the results on a scalable distributed file system such as the Hadoop file system (White, 2012), or by using a specialized distributed spatial data organization/retrieval system (Aji *et al.*, 2013; Hongchao and Wang, 2011). The master initializes the work by loading the tile map and assigning each slave to process a unique tile via a process tile (PT) message carrying the associated tile ID. Upon receiving a PT message, a slave loads and segments the tile and identifies the boundary data inside the tile consisting of eight disjoint sets (four edges and four corners). The slave outputs the segmented non-boundary trees, notifies the master via a tile complete (TC) message carrying the boundary sets, and waits for the master for a new assignment. The master then updates the tile map and inspects all of the eight boundary sets it received from the slave to determine if any of the associated edge/corner data is ready to be unified. Edge data is ready when both tiles sharing the edge are segmented and corner data is ready when all four tiles sharing the corner are segmented. The master then unifies all edge/corner data that are ready and re-assigns the waiting slave to re-segment the unified boundary data, which is conveyed by a process boundary (PB) message to the slave. The slave process, upon receiving the PB message, segments the boundary data conveyed by the message, outputs the result trees, and notifies the master via a boundary complete (BC) message. The master then re-assigns a new tile (chosen on an arbitrary order) via a PT message to the slave. If the master cannot locate any ready boundary data of the tile when it receives the TC message, it proceeds with re-assigning the waiting slave to segment a new tile via a PT message. If all tiles are segmented, the master terminates the slave process by sending a finalize (FIN) message. The master process continues until all slaves are finalized, implying that all tiles and their boundary data were processed.

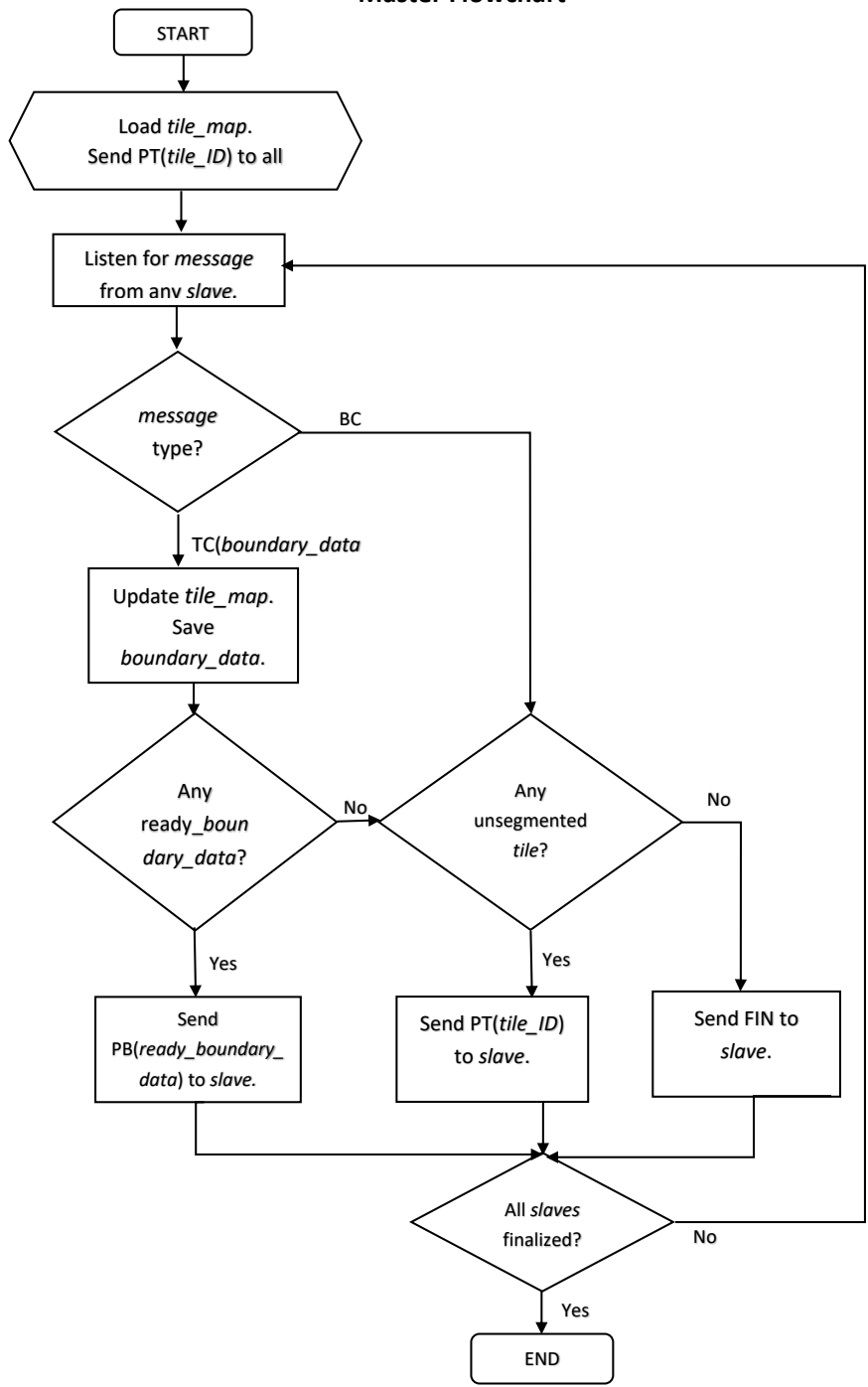

Figure 3. Flowchart of the master responsible for maintaining the tile map globally and coordinating the slaves.

**Slave Flowchart**

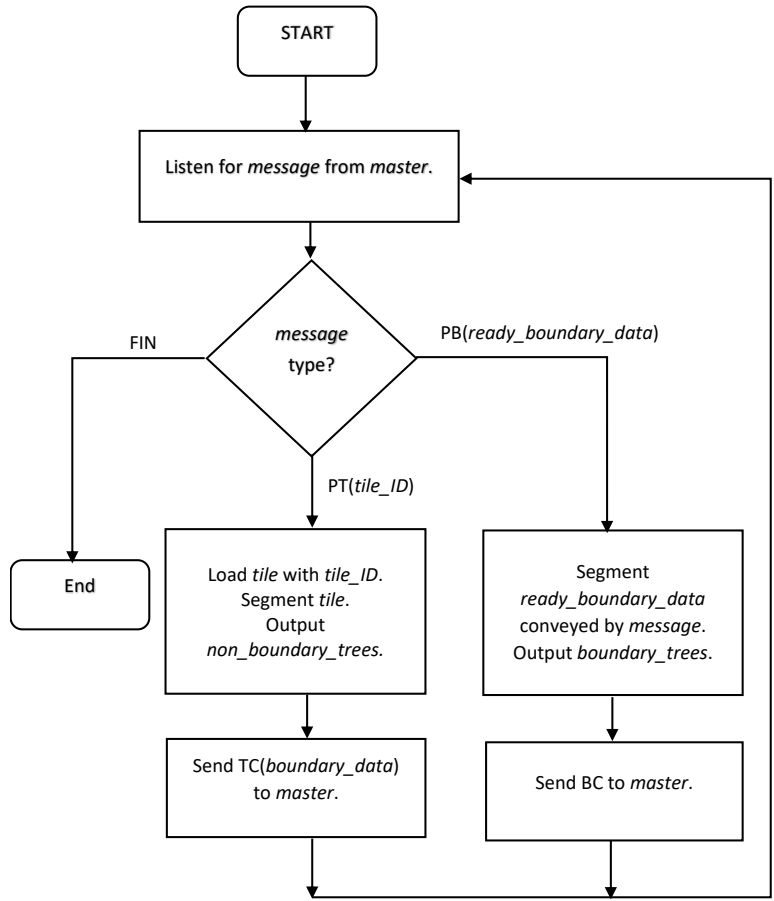

Figure 4. Flowchart of a slave segmenting tiles and boundary data as directed by the master.

In the presented distributed approach, all tile boundaries are guaranteed to be processed; once all tiles sharing each specific edge or corner are segmented, the edge/corner data is assigned to be processed by the slave that completed the last tile. Also, assuming that the amount of processing incurred by the master does not affect its responsiveness (theoretical limits are derived in the next section), the slaves keep working all the time resulting in an efficient distributed processing scheme.

## 2.3 Theoretical analysis

We assume that the entire LiDAR data consists of *N* points, which is arranged in tiles of *n* points on average, and LiDAR data representing each tree consist of *t* points ($n \gg t$). We assume that the single-processor tile segmentation algorithm has an asymptotic runtime complexity of $T_s(n)$. To illustrate, we assume that *p* processors can be allocated for processing *N/n* tiles ($N/n > p$).

The number of trees within a tile is proportional to the area of the tile while the number of trees along a tile edge is proportional to the edge length. Hence, given the average number of trees within a tile is *n/t*, the number of trees along one edge of the tile is the square root of it ($n^{1/2}/t^{1/2}$). Multiplying the number of trees along the edge by *t* results in $t^{1/2}.n^{1/2}$ LiDAR points per edge data. Therefore, the asymptotic runtime of re-segmenting the boundary data of a tile is $T_s(n^{1/2}.t^{1/2})$. Also, the communication of the boundary data between the master and a slave takes $O(n^{1/2}.t^{1/2})$. Each slave also needs to wait for the master to receive its boundary data, update its internal tile map, and re-assign the slave. Assuming the responsiveness of the master, this wait time is also bounded by $O(n^{1/2}.t^{1/2})$ because the master processes all of the LiDAR points it communicates with the slave. Aggregating the required time for re-segmenting, communicating data, and waiting for the master, the overhead for processing the boundary data is $T_s(n^{1/2}.t^{1/2}) + O(n^{1/2}.t^{1/2})$. Therefore, the efficiency of a single slave when segmenting a tile in the distributed approach presented above is given by:

$$e_s = \frac{T_s(n)}{T_s(n) + T_s(n^{1/2}.t^{1/2}) + O(n^{1/2}.t^{1/2})} \quad\quad 1$$

where $e_s$ denotes the efficiency of the slave; the numerator is the effective work; and the denominator is the total work including the effective work and the overhead.

Because the master does not perform segmentation, the entire segmentation that is performed by all of the *p-1* slaves is sped up by a factor of *(p-1)e_s*. Between the time when the first and the last slaves are finalized, the remaining workload of each active slave is bounded by *n* LiDAR points because each of them has at most one tile to complete. As soon as the first slave is finalized, a non-parallelizable workload is introduced to the distributed scheme. Between the time the first and the second slaves are finalized, the active slaves process with a missing fraction of the entire slaves' power, i.e., *1/(p-1)* of the power was already finalized. This results in *n/(p-1)* non-parallelizable workload. Similarly, between the time the *(i-1)$^{th}$* and *i$^{th}$* slaves are finalized, *(i-1)n/(p-1)* non-parallelizable workload is introduced. Therefore, the total non-parallelizable workload is:

$$w_s = \sum_{i=2}^{p-1} \frac{i-1}{p-1} n = \frac{p-2}{2} n \qquad 2$$

where $w_s$ denotes the non-parallelizable (serial) workload of the entire distributed processing (the initialization workload performed by the master is a negligible constant. Hence, the ratio (*P*) of the parallelizable (total minus serial) workload to the total workload is:

$$P = \frac{N - \frac{p-2}{2} n}{N} \qquad 3$$

Finally, the speedup of the entire distributed approach denoted by $S_p$ according to Gustafson-Barsis law (McCool *et al.*, 2012) is:

$$S_p = 1 - P + P(p-1)e_s \qquad 4$$

The time the master requires to devote per tile is proportional to the number of LiDAR points it deals with, which is $O(n^{1/2}.t^{1/2})$, while the time a slave requires to devote per tile is $T_s(n) + T_s(n^{1/2}.t^{1/2}) +$ $O(n^{1/2}.t^{1/2})$. Thus, in order for the master to remain responsive for *p-1* slaves so that the above equations hold, we should have:

$$p-1 \leq \frac{T_s(n)+T_s(n^{1/2}.t^{1/2})+O(n^{1/2}.t^{1/2})}{O(n^{1/2}.t^{1/2})} \qquad 5$$

## 3 Results and discussions

### 3.1 Runtime and scalability

We adopted the tree segmentation algorithm presented by Hamraz et al. (2016) as the single-processor building block to empirically assess the proposed distributed processing approach. The tree segmentation algorithm can efficiently be implemented such that $T_s(n) = O(n)$ (see Appendix A). We implemented the master-slave scheme using the MPI and ran it on the University of Kentucky Lipscomb cluster, which has 256 symmetric basic nodes (Dell C6220 Server, 4 nodes per 2U chassis), each with 16 cores (dual Intel E5-2670 8 Core – Sandy Bridge) at 2.6 GHz and 64 GB of RAM at 1,600 MHz. The nodes are interconnected via Mellanox Fourteen Data Rate InfiniBand (2:1 over-subscription, 14.0625 Gbit/s) and equipped with a global file system (DDN GridScaler SFA12K storage appliance with the IBM GPFS – Read: 25 GB/s throughput and 780,000 IO/S, Write: 22 GB/s throughput and 690,000 IO/S) (University of Kentucky Analytics & Technologies). We experimented with four contiguous loads of data: the first 200 (Figure 1 – counting row-wise starting from the top leftmost tile toward right and then down), 400, and 600 tiles, as well as all 801 tiles. For each load, we ran the distributed segmentation approach using 1–12 computing nodes (i.e., 16, 32, …, 192 processing cores), and measured the experimental speedups by dividing the observed single-processor runtime by the observed distributed processing runtimes. The observed single-processor runtime equals the number of tiles multiplied by average observed runtime of a tile, which equaled 31 minutes and 8 seconds (2.8% loading from disk, 94.8% computation, and 2.4% writing to disk) averaged for a sample of 128 tiles.

Figure 5 shows the experimental speedups overlaying the equivalent theoretical speedups using Equation 4 for which *t = 1,350* and *n = 5×10$^6$* as measured in the dataset. In order to calculate the exact value of $e_s$ using Equation 1, the constant coefficients of the asymptotic functions in the numerator and the denominator need to be measured on the specific runtime platform. According to our measurement, the ratio of the constant coefficient of the numerator ($T_s(n)$ – equals to $O(n)$ here) to the constant coefficient of $O(n^{1/2}.t^{1/2})$ appeared in the denominator is about *150*. In other words, the time required for the segmentation of a LiDAR point cloud is approximately 150 times greater than the time required for two-way inter-process communication (from a slave to the master and back) of the same size point cloud on our runtime platform. Substituting the values of *t*, *n*, and the ratio of the constant coefficients in Equation 1, the efficiency of a slave ($e_s$) equals *0.9837*. Similarly, using Equation 5, having *p-1 ≤ 9,279* renders the master to remain responsive.

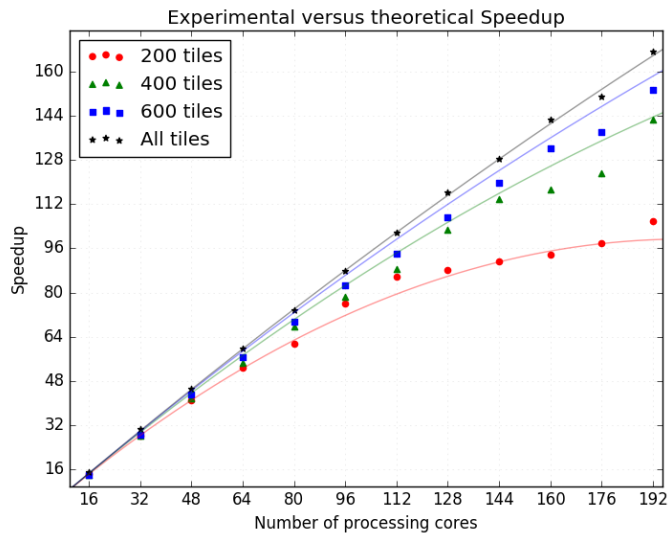

Figure 5 Experimental speedups shown by symbols, which overlay corresponding continuously drawn theoretical speedups for different loads of data.

As shown in Figure 5, processing the entire tiles using 192 processing cores resulted in a practical speedup of 167.04 (compared to 165.70 of theory), meaning that we reduced the expected single-processor runtime of over 17 days to 2 hours and 29 minutes. Although a few weeks of processing time might be acceptable for forest inventory to be performed annually, it is infeasible for potential real-time applications, e.g., more accurate aerial monitoring of wildfire using LiDAR (Arroyo *et al.*, 2008; Contreras, 2010). After all, natural forests may be several times greater than Robinson Forest and be recorded with greater point densities (to become affordable given the advancements of the sensor technology) yielding much larger datasets, which even more justifies the need for distributed processing.

The small differences between the empirical and the theoretical speedups (Figure 5) are likely due to natural variabilities in the dataset as well as small differences in the runtime environment from the theoretical assumptions. These results show that the distributed segmentation approach can achieve nearly linear speedup using a reasonable number of processing cores and given a sufficiently large dataset (at least two times more tiles than the number of cores). Because the number of tiles is typically large for forest-level data and the number of cores is limited, scalability of the approach to arbitrarily large datasets is fulfilled.

As the distributed approach does not assume a fixed number of slave processes, it can also be implemented on a grid environment in which the master can be in charge of initiating new slave processes and rescheduling tasks in case of node failure. In case Equation 5 is violated (the master is overloaded), the straightforward solution is to increase the size of tiles to make the slaves perform proportionately more work per each tile assignment. A more flexible solution is to augment the distributed scheme to accommodate multiple masters in a hierarchical fashion. An additional improvement might consider slaves not sending boundary data to the master. Instead, they can set aside the data in a buffer and send it later on directly to the slave who would eventually process the boundary data. In this case, the master should be in charge of coordinating the interactions between the slaves and would not need to deal with receiving and sending boundary data, which decreases the master's workload and make it independent of the tile size. Such an improvement would not affect the asymptotic calculations of speedup presented above, even though it may help to reduce the runtime in practice specially if the master is overloaded and/or the inter-process communication on the runtime platform is costly. Lastly, the master can employ any strategy for choosing a new tile to assign next without affecting the final result and the processing time in theory, although assigning contiguous tiles makes boundary data become ready earlier and results in freeing up memory earlier, which may become invaluable depending on the circumstances.

Tailoring the proposed approach to run under the Hadoop MapReduce infrastructure in a single stage can also be accomplished as follows. Loading and segmentation of an individual tile should be defined as the map phase, in which the non-boundary trees should be output to the file system and each of the eight boundary data are assigned a unique key for the reduce phase. The unique key of each specific edge/corner data should be the same across all the map tasks that share the specific edge/corner. The reduce phase should be defined to unify all of the data it is given (edge/corner data portions having an identical unique key), re-segment the data, and output the result to the file system. There would not be an explicitly defined master process because the underlying map-reduce infrastructure is responsible for coordination between the map and the reduce tasks, as well as scalability and fault tolerance of the entire ecosystem. In contrast, the MPI implementation using a global scalable file system generally runs faster because slaves barely idle, while reduce phase cannot start processing until map phase finishes. This performance advantage is achieved because of having explicit control over the inter-process communications enabling design of a flexible scheduling scheme using MPI, although it generally requires more effort and expertise to design and program desired features for a distributed application.

## 3.2 Global forest parameters

Although tile size does not affect the segmentation result of the distributed approach in theory, depending on the underlying single-processor segmentation algorithm, it may introduce slight biases in practice. Such biases have a direct correlation with the total length of the shared edges of the tiles because the boundary data along those edges are indeed the only places that are not processed exactly the same compared to a single-processor run. In order to quantify the biases in terms of number of trees, we processed five sample square (1.524 Km side ~ 232.5 ha area) blocks (each composed of 5×5 tiles) in a single-processor manner as well as using the distributed approach. We partitioned each block to uniform grids of 2×2, 3×3, …, 15×15 sub-blocks and ran the distributed approach for each of the grid patterns. Single-processor execution detected an average of 62,005 trees in a block. Figure 6 shows the average number of trees detected per block as a function of the total length of the shared edges of sub-blocks, which equals $2 \times (n_{sb}-1)$ multiplied by the block side length where $n_{sb}$ denotes number of sub-blocks along a block side. As expected, additional number of trees compared with single-processor run shows a linear relation with the total shared edge length: an average of 96 additional trees (false positives) were detected per 1 Km of shared edge, which is a small value given that more than 26,000 trees were detected per 1 $Km^2$.

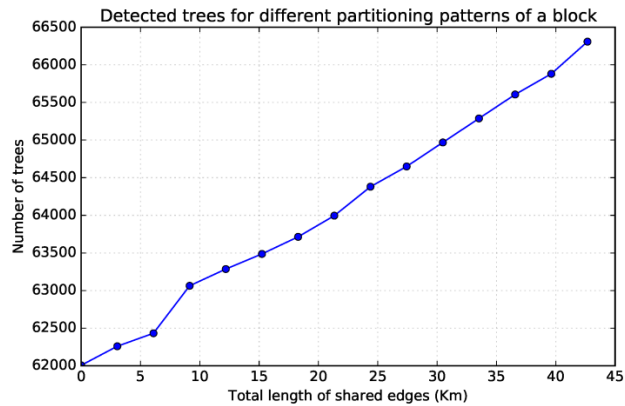

Figure 6. Number of trees detected in a block for different partitioning patterns.

When applied to the entire Robinson Forest, the distributed tree segmentation approach detected a grand total of 1,994,970 trees over the area covered by the LiDAR data. The total length of shared edges in the tile map (Figure 1) is 446.23 Km, which results in 42,833 potential false positives (2.15%) be introduced across the tile edges. When the number of false positives is subtracted, the grand total of detected trees becomes 1,952,137.

Due to imperfectness of the single-processor algorithm, a portion of grand total number of detected trees was associated with over-segmentations, and a portion of existing trees in the forest was undetected. In order to account for the over-segmentations/undetected trees, we used the accuracy result of the single-processor segmentation algorithm on the LiDAR point clouds (taken from the same dataset) of a field-surveyed sample of 23 (0.1 ac) circular plots placed across Robinson Forest (Hamraz *et al.*, 2016). The accuracy result included the number of detected trees (bearing over-segmentations) and the number of existing trees (bearing undetected trees) per four crown classes (dominant, co-dominant, intermediate, and overtopped). Within each of the 23 plots, we calculated a fraction per crown class: the existing trees of that crown class divided by the grand total (all crown classes) of detected trees. Table 1 shows the mean and 95% T-confidence bounds of the fractions across the 23 plots. It also shows the adjusted estimates of number of existing trees, which were calculated by multiplying the grand total number of detected trees using the distributed approach to the corresponding fractions. Considering a 95% T-confidence interval, the total number of existing trees in the 7,441.5-ha forested area is estimated to be 2,495,170 (±13.52%), which results in an average of 335.30 trees per ha.

Table 1 Estimated number of trees categorized based on tree crown class.

| Crown Class | Fraction of existing to grand total detected | | Estimated number of existing trees | |
|---|---|---|---|---|
| | mean | 95TCB[1] | entire forest | per ha |
| Dominant | 0.0785 | ±75.50% | 153,178 | 20.59 |
| Co-dominant | 0.3069 | ±23.07% | 599,106 | 80.50 |
| Intermediate | 0.5376 | ±17.84% | 1,049,446 | 141.32 |
| Overtopped | 0.2928 | ±43.29% | 571,522 | 76.80 |
| Dead | 0.0625 | ±104.7% | 121,917 | 16.38 |
| All | 1.2782 | ±13.52% | 2,495,170 | 335.30 |

[1] 95% T-Confidence Bounds (DF=22)

For verification, we compared our tree number estimates (Table 1) with equivalent estimates based on field measurements of another sample of 23 plots from the Robinson Forest (Figure 7). The estimates for total number of trees differ by about 3% and the estimates of number of dominant trees differ by about 30%. However, the large overlap between the 95% T-confidence interval errors indicate that there is no statistically significant difference between the estimates using LiDAR and the field collected measurements.

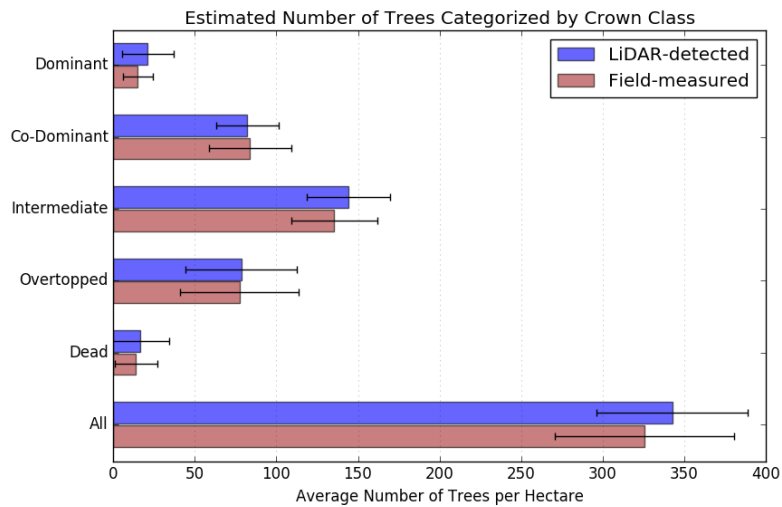

Figure 7. Estimated number of trees using LiDAR compared to field-collected along with the 95% T-confidence intervals.

Figure 8 shows the height distribution of all detected trees (heights above 5 m) by the approach. The height distribution follows a bimodal pattern, which can be attributed to multistory structure of deciduous natural forests, in which the dominant and co-dominant trees form the over-story and intermediate and overtopped trees form the mid-story. We fitted a normal mixture model to the bimodal distribution: the larger lump on the right (associated with over-story trees) has a mean height of 26.9 m and a standard deviation of 6.6 m, and the smaller lump (associated with mid-story trees) has a mean height of 9.4 m and a standard deviation of 2.6 m.

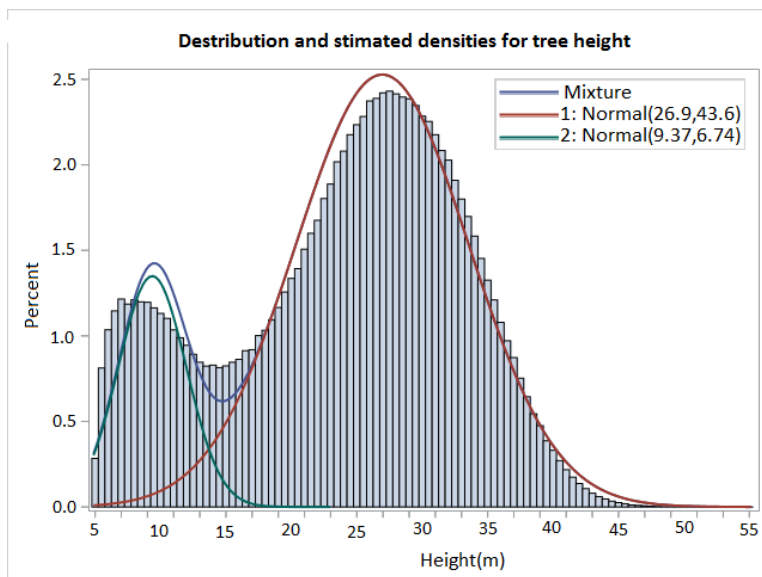

Figure 8 Height distribution of 1,994,970 trees detected in Robinson Forest superimposed with estimated normal mixture model.

We compared the LiDAR-derived mean tree height estimates with those obtained using a field sample from the forest (371 over-story and 826 mid-story trees). The sample mean height of the over-story trees was 25.4 m with a standard deviation of 5.3 m, and the sample mean height of mid-story trees was 17.0 m with a standard deviation of 4.1 m. Considering that the LiDAR-detected tree heights are in fact biased by presence of falsely detected trees and absence of undetected trees, the field estimates are close to the LiDAR-detected estimates for over-story trees. However, the field estimates for mid-story trees are remarkably larger than the LiDAR-detected estimates, which can be justified as follows. Airborne LiDAR provides considerably less information about the mid-story trees due to decreased penetration of LiDAR points toward bottom canopy layers (Maguya *et al.*, 2014; Reutebuch *et al.*, 2003), hence detected tree rate is lower for mid-story trees (Duncanson *et al.*, 2014; Hamraz *et al.*, 2016). Also, the detected mid-story trees are likely biased to be smaller within the population of all existing mid-story trees because they are easier to detect when there is less canopy closure, which is associated with stand age and is minimal when stand is young and in general has smaller trees (Jules *et al.*, 2008). So, detecting relatively fewer mid-story trees that are also likely biased to be smaller leads to capturing a distribution with smaller mean and standard deviation. After all, the only information used to fit the normal mixture model was the heights of the trees while height may not be sufficient for classification, i.e., a moderately tall tree can in fact be mid-story if situated in a taller stand while the mixture model always probabilizes it strongly as over-story according to its height, and vice versa. Thus, the procedure of fitting the normal mixture model likely separates the two tree classes more distantly with respect to height.

In addition to providing number of trees and height distributions (compared here to field surveys for validation), the distributed approach enables identification of individual tree locations and attributes (tree height and crown widths) as well as the point cloud segments representing tree crowns for large forested areas in a timely manner, which in turn enables building a detailed (at the individual tree level) forest model and performing a myriad of more accurate analyses. For instance, tree attributes can be used to develop allometric equations to estimate other important tree metrics such as DBH and volume, and the point cloud segments can be used to construct the 3D geometric shape of each individual tree crown to develop mode detail estimates such as crown volume, biomass, and carbon content.

### 3.3 Approach application to other spatial datasets

As mentioned earlier, the approach uses a single-processor tree segmentation algorithm as a building block and does not require any knowledge on how the algorithm functions. So, the approach may be used to straightforwardly adopt any other single-processor object identification/segmentation algorithm in order to scale up processing arbitrarily big spatial and geospatial datasets, such as remotely sensed buildings, cars, planets, etc. The only caveat is that the objects may not be greater than the tiles, i.e., they may not touch more than two adjacent edges of a tile.

Moreover, generalization of the approach to process 3D spatial data can be accomplished similarly as follows. Instead of tiles that are representing surfaces, cubes representing volumes will be the data units for 3D data. Boundary data in this case would be surface (shared between two cubes), edge (shared among four cubes, and corner (shared among eight cubes) that can be handled for distributed processing using the master-slave processing scheme presented in Section 2.2. The theoretical runtime analysis for 3D data would be slightly different. The average number of the entire objects within the cube is proportional to the cube volume while the number of boundary objects (those touching a cube surface) is proportional to the cube surface area. Hence, the number of boundary objects equals the number of objects within the cube raised to *2/3* power, which changes the master/slave overheads and Equations 1 and 5 need to be updated accordingly.

## 4 Conclusions

Obtaining tree-level information over large forested areas is increasingly important for accurate assessment, monitoring and management of forests and natural resources. Several automated tree segmentation methods have been developed, but these methods have only been applied to small forested areas for accuracy assessment. Although these methods can in theory be applied to larger areas, such applications is not straightforward because LiDAR data covering forest-level data far exceeds the memory of desktop computers and may also take unacceptably long time to be processed sequentially.

Here we presented and analyzed a scalable distributed approach that was applied to segment trees within a LiDAR point cloud covering an entire forest. The distributed approach segmented trees within the tiles and across the tile boundaries, and introduced a minimal bias compared with the single-processor algorithm that was also quantified in this work. Comparison of the estimated number of trees and the tree height distribution with the field surveys validated sound operation of the approach. We presented the distributed processing approach and the associated analysis in a platform-independent manner so as the implementation can be accomplished on different distributed platforms with minor modifications.

The presented approach enables obtaining individual tree locations and point cloud segments representing the tree crowns for entire forested areas in a timely manner. The resulting detailed, tree-level information has the potential to increase the accuracy of forest level information by creating remotely sensed forest inventories for more efficient management of forest and natural resources. Although the distributed approach was presented within the context of tree segmentation from LiDAR point clouds, it can straightforwardly be applied to segment/identify objects within other large-scale datasets.

## Acknowledgments

This work was supported by: 1) the University of Kentucky Forestry Department and the McIntire-Stennis project KY009026 Accession 1001477, ii) the Kentucky Science and Engineering Foundation under the grant KSEF-3405-RDE-018, and iii) the University of Kentucky Center for Computational Sciences.

# Appendix A: Efficient implementation of segmentation algorithm and runtime analysis

The tree segmentation algorithm we used as a building block in this study (Hamraz *et al.*, 2016) consists of a pre-processing step including homogenizing the point cloud, removing non-surface points, smoothing, and then a loop over the five major steps outlined below until the entire point cloud is clustered: 1) locate the non-clustered highest point - global maximum (GMX); 2) generate vertical profiles originating from the GMX with a length of maximum tree crown radius; 3) For each profile, identify the LiDAR point along the profile that represents the crown boundary; 4) create a convex hull of the identified boundary points; and 5) cluster all LiDAR points encompassed within the convex hull as the highest tree crown.

For an efficient implementation, the point cloud should be indexed in a 2D horizontal grid. Indexing and the pre-processing step takes *O(n)* where *n* is the number of points. We assume that the main loop iterates *m* times. Naively locating the GMX (step 1) takes *O(n) per iteration*. Instead, we create a descendingly sorted list of all of the grid cells according to the height of the point they contain and mark all cells as unvisited. The sorting procedure takes $O(n.\log n)$. The grid cells are marked as visited when they are clustered in step 5. To locate the non-clustered GMX, the sorted list is traversed from the position of the previous GMX forward, which on average takes *O(n/m)* per iteration. Once the GMX is located, clustering the highest tree (steps 2–5) has a runtime independent of *n* and *m* and is proportional to the tree size, which is bounded and can be assumed as a constant. So, the aggregate runtime of each iteration of the loop is *O(m/n)*, hence the total runtime of the loop becomes O*(n)*. Aggregating the pre-processing and the sorting times:

$$T_s(n) = O(n) + O(n.\log n) \qquad \text{A.1}$$

where $T_s(n)$ is the total runtime of the algorithm; the first term on the right-hand side corresponds to the runtime of the main loop and the pre-processing step; and the second term corresponds to the runtime of the sorting procedure before the loop.

We ran the implementation above on a workstation of 3.4 GHz CPU speed and 8 GB of RAM for 25 loads of data. Figure A.1 shows the log-log plot of the runtime of the segmentation versus the number of points.

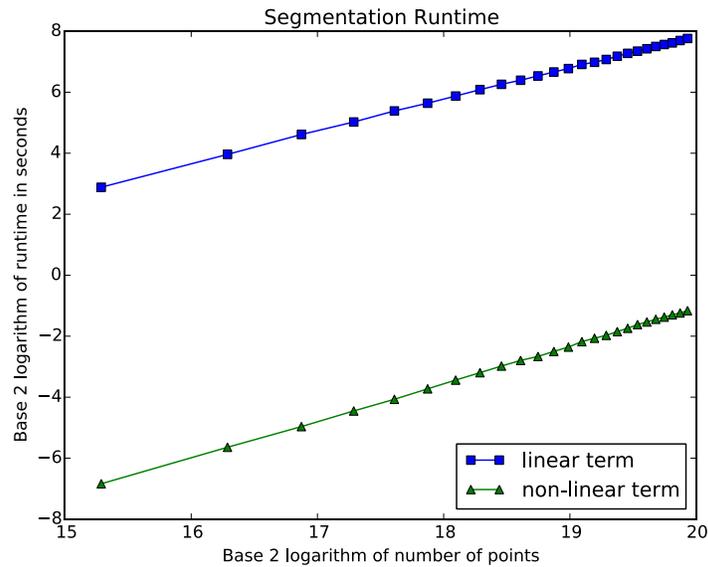

**Figure A.1.** Log-log plot of the Segmentation runtime versus the number of LiDAR points in the point cloud. Each symbol corresponds to average across 15 strata.

The slope of the best fit line to the square symbols is *1.03*, which concurs with the linear term of Equation A.1. Also, the triangle symbols show a slightly super-linear pattern concurring with the non-linear term. We measured the constant coefficients of both terms by dividing the execution times associated with the terms by $n$ and $n \cdot \log n$ respectively. The ratio of the linear coefficient to the non-linear one is platform-independent and is about *7,800* according to our measurement. This yields that $n$ should be greater than $2^{7,800}$ in order for the non-linear term to start dominating the linear term, which corresponds to a LiDAR point cloud covering over *3e+2,331* times surface area of the earth. So, we may safely replace $\log n$ in the

non-linear term with an upper bound constant, which reduces Equation A.1 to:

$$T_s(n) = O(n) \qquad \text{A.2}$$